\documentclass[twocolumn,superscriptaddress,showpacs,preprintnumbers,amsmath,amssymb,prl]{revtex4-1}

\usepackage{verbatim}
\usepackage{epsfig}
\usepackage{graphicx}

\usepackage{color}
\usepackage[colorlinks,bookmarks=false,citecolor=blue,linkcolor=red,urlcolor=blue]{hyperref}
\usepackage{dcolumn}
\usepackage{bm}

\begin{document}

\title{Vortex core order and field-driven phase coexistence in the attractive Hubbard model}

\author{Madhuparna Karmakar}
\email{madhuparna.k@gmail.com}
\affiliation{The Institute of Mathematical Sciences, HBNI, C I T Campus, Chennai 600 113, India}
\author{Gautam I. Menon}
\email{menon@imsc.res.in}
\affiliation{The Institute of Mathematical Sciences, HBNI, C I T Campus, Chennai 600 113, India}
\author{R. Ganesh}
\email{ganesh@imsc.res.in}
\affiliation{The Institute of Mathematical Sciences, HBNI, C I T Campus, Chennai 600 113, India}
\date{\today}

\begin{abstract}
Superconductivity occurs in the proximity of other competing orders in
a wide variety of materials. Such competing phases may reveal themselves when superconductivity is locally suppressed by a magnetic field in the core of a vortex. We explore
the competition between superconductivity and charge density wave order in the attractive Hubbard model on a square lattice.
Using Bogoliubov-deGennes mean field theory, 
we study how vortex structures form and evolve as the magnetic flux is tuned. 
Each vortex seeds a CDW region whose size is determined by the energy cost of the competing phase.
The vortices form a lattice whose lattice parameter shrinks with increasing flux. Eventually, 
their charge-ordered vortex cores overlap, leading to a
field-driven coexistence phase exhibiting  both macroscopic charge order and superconductivity -- a `supersolid'. 
Ultimately, superconductivity disappears via a first-order phase transition into a purely charge ordered state. 
We construct a phase diagram containing these multiple ordered states, using $t'$, the next-nearest neighbour hopping, to tune the competition between phases.
 
\end{abstract}
\pacs{} 
\keywords{}
\maketitle
\paragraph{Introduction:}
Superconductivity is often obtained in proximity to other ordered ground states. The most prominent example being the high T$_c$ cuprates, where superconductivity competes with antiferromagnetism and with charge order\cite{Lake2001,Chang2012}. A particularly interesting way to stabilize underlying competing phases is to apply a magnetic field, locally suppressing superconductivity to create vortices. The core region of the vortex can then host competing correlations\cite{vortexcoreAFM,Curran2011,Machida2015}. Indeed, experiments with scanning tunnelling microscopy have revealed charge-ordered\cite{vortexcoreAFM,Machida2015} vortex cores in the cuprates. NMR studies of YBa$_2$Cu$_3$O$_y$ indicate that as the magnetic field increases, the inter-vortex distance decreases; at a critical field strength, vortex cores overlap leading to  charge order throughout the system\cite{Wu2013}. 
These and related experiments motivate the study of vortex core order and field-driven coexistence in the attractive Hubbard model, the simplest model to show competition between superconductivity (SC) and charge density wave (CDW) order.

\begin{figure}
\includegraphics[width=3.5in]{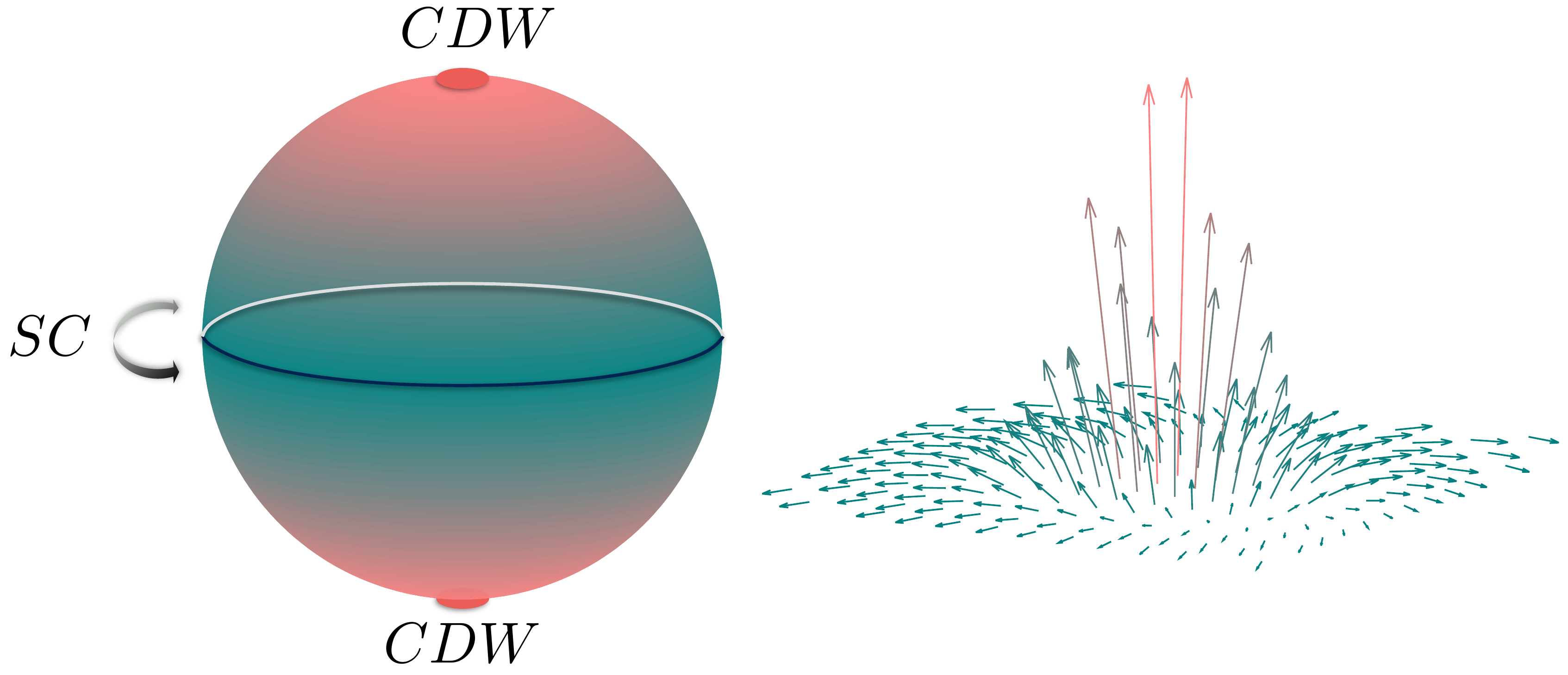} 
\caption{Left: The space of order parameters forming an $SO(3)$ sphere; the equator corresponds to the $U(1)$ phase of the SC order parameter while the poles correspond to two possible checkerboard CDW orders. A generic point on the sphere represents coexisting SC and CDW orders.
Right: The order parameters forming a `meron' in the vicinity of a vortex, with $t'=0.3t$. Far from the core, the pseudospins lie in the plane and wind by $2\pi$ as we move around the vortex. Within the core, they cant out of the plane to generate CDW order.  
}
\label{fig.SO3}
\end{figure}

\paragraph{Hubbard model and $SO(3)$ symmetry:} 
We consider fermions on a square lattice, described by 
\begin{equation}
 H \!=\! \sum_{\langle ij \rangle,\sigma} \left\{- t_{ij} c_{i,\sigma}^\dagger c_{j,\sigma} +
 h.c.\right\} -U\sum_{i}\hat{n}_{i,\uparrow} \hat{n}_{i,\downarrow} - \mu \sum_{i,\sigma} \hat{n}_{i,\sigma}.
 \label{eq.Hamiltonian}
 \end{equation}
\noindent where $\mu$ is the chemical potential and $U$ is the strength of the on-site attractive interaction ($U > 0$). The hopping parameter $t_{ij}$ takes the value $t$ (henceforth set to unity) for nearest neighbours and is zero otherwise.
When $\mu$ is tuned to half-filling, this model possesses a remarkable symmetry with SC and CDW order becoming degenerate; the order parameters form an enlarged space having $SO(3)$ symmetry as shown in Fig.~\ref{fig.SO3}(left)\cite{Yang1989,Zhang1990,Yang1990,Burkov2008,Ganesh2009}. This is a delicate symmetry arising from the bipartite nature of the square lattice with hoppings connecting sites of different sublattices. We can tune away from this $SO(3)$ degenerate point by introducing a next-nearest neighbour hopping, $t'$. The $t'$ term lowers the energy of the SC phase relative to CDW phase. 

The $SO(3)$ degeneracy leads to a local pseudospin order parameter whose components are $\left\{ \mathrm{Real}(\Delta_i),\mathrm{Imag}(\Delta_i),\tilde{\phi}_i\right\}$ as shown in Fig.~\ref{fig.SO3}. Here, $\Delta_i$ and $\tilde{\phi}_i$ (defined below) are the local superconducting and CDW order parameters. This $SO(3)$ symmetry is directly analogous to the hypothesized $SO(5)$ symmetry\cite{Demler2004} in the cuprates which groups SC and antiferromagnetism into an enlarged order parameter space. As a testable consequence of $SO(5)$ theory, it was proposed that vortex cores would have antiferromagnetic order\cite{Arovas1997,Hu2002}. 
Analogously, the Hubbard model in Eq.~\ref{eq.Hamiltonian} will possess CDW order in the vortex core. 
In the language of $SO(3)$ pseudospins, a vortex corresponds to a `meron', as shown in Fig.~\ref{fig.SO3}(right) -- in the core region, the moments cant out of the plane to locally give rise to CDW order. Here, unlike in the cuprates, we have direct control over the $SO(3)$ symmetry breaking in the form of the $t'$ hopping. 
We study the Hubbard model in an applied field demonstrating CDW order in the vortex core.
 Our key result is a field-driven SC-CDW coexistence regime which arises from the overlap of vortex cores. This state simultaneously breaks translational symmetry and $U(1)$-gauge symmetry, demonstrating a new route to `supersolidity'\cite{Boninsegni2012}.

\begin{figure}
 \centerline{
 \includegraphics[height=4cm,width=4cm,angle=0]{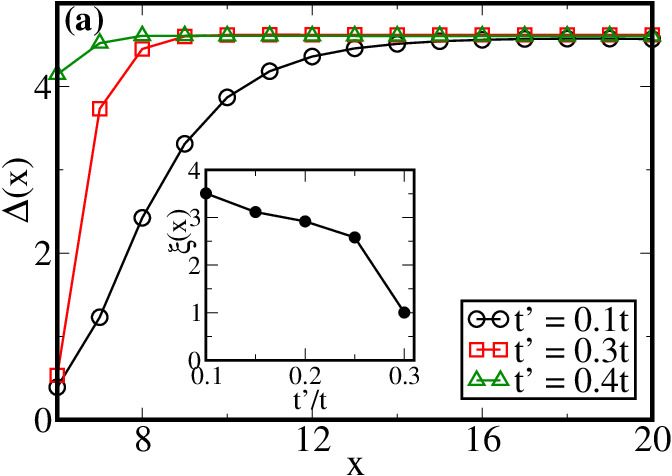}
 \includegraphics[height=4cm,width=4cm,angle=0]{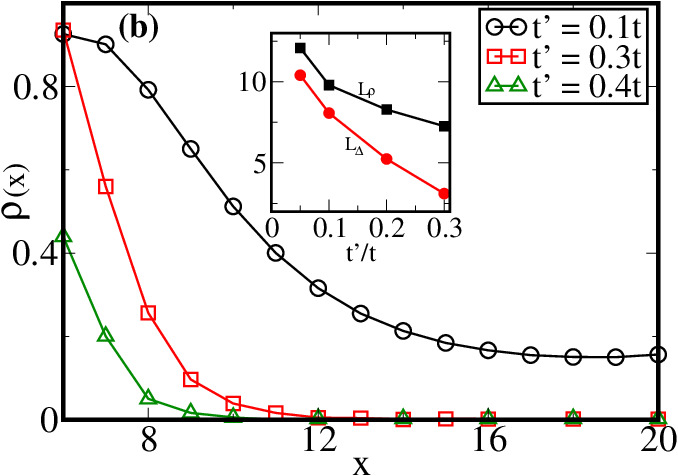}
 }
 \vspace{-0.1cm}
 \includegraphics[height=4cm,width=9cm,angle=0]{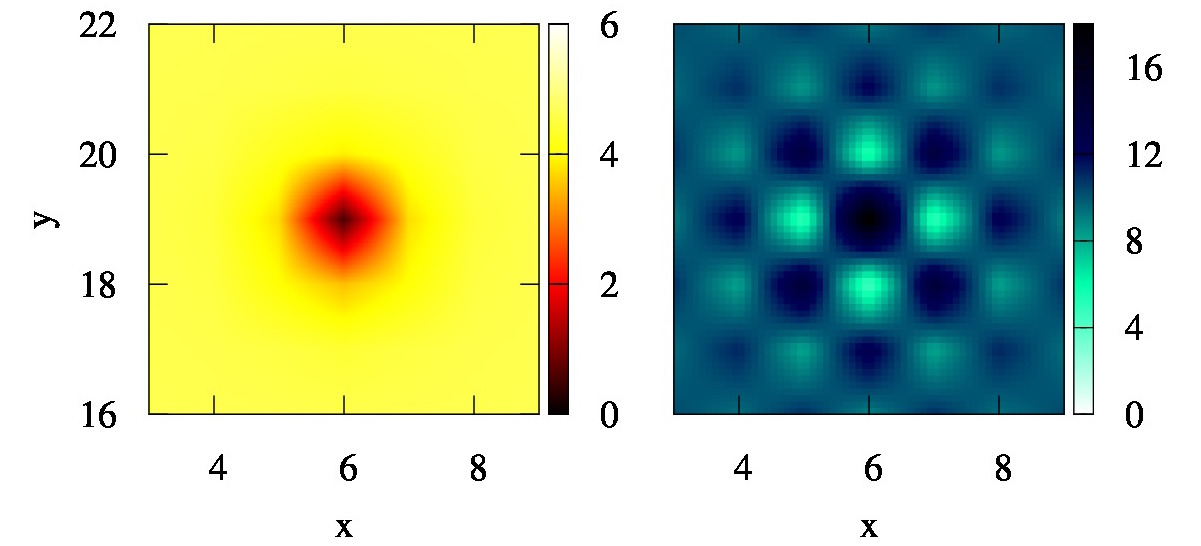}
 \caption{(a) Superconducting and (b) CDW order profiles at different 
 $t^{\prime}$. The inset to panel (a) shows the underlying length scale $\xi_\Delta$ vs. $t'$.
 The inset to panel (b) shows the FWHM widths, $L_\Delta$ and $L_\rho$, vs. $t'$.
 The lower panels show the spatial maps of SC($\vert \Delta_i \vert$) (left) and density($\phi_i$)(right) order parameters around a vortex core for 
 $t^{\prime} = 0.3t$. The interaction strength is fixed at 
 $U=10t$.}
\end{figure}

\paragraph{Bogoliubov deGennes mean-field theory:}
We perform simulations on an $L\times L$ lattice with periodic boundary conditions, with $L$ up to $30$. 
To introduce an orbital magnetic field, we add a complex phase to the hopping amplitudes $t_{ij}$ given by $\theta_{ij} = e\int_{\bf r_{i}}^{\bf r_{j}} {\bf A. dr}$, where ${\bf A}({\bf r})$ is the vector potential, see Supplementary Materials for details. 
The net magnetic flux through a closed surface must be quantized in units of $h/e$\cite{Dirac1931}. We take the net flux through our system to be $\alpha h/e$ where $\alpha$ is an integer. As each vortex carries a flux $\Phi_0=h/2e$, we will always have an even number of vortices in the system. In particular, the lowest magnetic flux we can have is $2\Phi_0$, corresponding to two vortices. 

We decompose the on-site interaction term in pairing and density channels. The SC order parameter is complex-valued, defined as $\Delta_{i} = U\langle c_{i\downarrow}c_{i\uparrow}\rangle$. The density order parameter is defined as $\phi_{i} = \frac{U}{2}\langle \hat{n}_{i\uparrow}+\hat{n}_{i\downarrow}\rangle = 
\frac{U}{2}(\langle c_{i\uparrow}^{\dagger}c_{i\uparrow}\rangle + \langle c_{i\downarrow}^{\dagger}c_{i\downarrow}\rangle)$. 
The local CDW order parameter can be defined as $\tilde{\phi}_i=(-1)^{\mathbf{r}_i}\{\phi_i-U/2\}$, which measures the local deviation from half-filling. 
With these mean-field parameters, the Hamiltonian takes the form of a $2L^2 \times 2L^2$ matrix, which can be diagonalized using the Bogoliubov-Valatin transformation \cite{ting2002, ting2004, han}.
We obtain self-consistent values of $\Delta_i$ and $\phi_i$ on every site. 

We find several self-consistent mean-field configurations, of which the one with lowest energy is to be chosen. One solution is a pure CDW state in which $\Delta_i = 0$ for all $i$ and $\phi_i = \left\{\phi_0 + (-1)^{\mathbf{r}_i} \tilde{\phi}\right\}$, corresponding to uniform CDW order. 
In the absence of a magnetic field and in the presence of a non-zero $t'$, 
this state has higher energy than the uniform SC phase. 
When a field is imposed, this state is not affected as it is insulating -- its energy remains constant, independent of the flux (see Supplementary Materials). In contrast, the SC phase necessarily develops vortices when a field is imposed. As the number of vortices increases with flux, so does the energy of the SC. As seen from these energetic arguments, an applied magnetic field induces competition between SC and CDW orders.

The $SO(3)$ symmetry of the attractive Hubbard model only exists precisely at half-filling. As we are interested in phase competition, all results presented here are at half-filling. We present results for $U=10t$ for the following reason. At large $U$, the Hubbard model can be mapped to a spin problem with antiferromagnetic superexchange interactions\cite{Burkov2008,Ganesh2009}. The local order parameter is, in fact, the $SO(3)$ spin whose components are $\left\{ \mathrm{Real}(\Delta_i),\mathrm{Imag}(\Delta_i),\tilde{\phi}_i\right\}$ as shown in Fig.~\ref{fig.SO3}. 
At low temperatures, we expect the system to have uniform spin length, ie., $\vert \Delta_i \vert^2 + \tilde{\phi}_i^2 = c$, a constant independent of position. SC and CDW order parameters are not independent, having to satisfy this uniform-length constraint\cite{Arovas1997}. With these considerations, the appropriate Landau Ginzburg free energy density is given by\cite{Arovas1997}
\begin{eqnarray}
\nonumber \mathcal{L} = \frac{\rho}{2} \left| \left( \mathbf{\nabla} - \frac{ie}{\hbar c}\mathbf{A} \right) \Delta(\mathbf{r} ) \right|^2
+ \frac{1}{8\pi} \left( \mathbf{\nabla} \times \mathbf{A} \right)^2 \\
+ \frac{\rho}{2} \vert \mathbf{\nabla}\tilde{\phi}(\mathbf{r} ) \vert^2
-\vert \Delta (\mathbf{r} )\vert^2 - (1- gt'^2)\vert \tilde{\phi} (\mathbf{r} )\vert^2.
\label{eq.LG}
\end{eqnarray}
The order parameters are coupled by the uniform length constraint: $\vert \Delta (\mathbf{r} ) \vert^2 + \tilde{\phi}^2 (\mathbf{r} )= c$. 
SC and CDW orders become degenerate when $t'=0$ and the magnetic field is turned off, revealing the underlying $SO(3)$ symmetry. 
At $U=10t$, we find that the mean-field results always satisfy the uniform spin length constraint and the above Landau Ginzburg theory applies. 
We find the same qualitative results extending to small $U$ values as well. 

Motivated by recent experiments revealing charge order in the cuprates, several authors have studied field theories similar to Eq.~\ref{eq.LG}\cite{Efetov2013,Hayward2014,Wachtel2014} with Ref.~\onlinecite{Meier2013} also incorporating an orbital magnetic field. Our study of the attractive Hubbard model at strong coupling can be viewed as an ultraviolet regularization of such a field theory.

\paragraph{Vortex profile:}
Setting $\alpha=1$, we obtain the lowest flux configuration with two well-separated vortices. 
As $t'$ is increased, we find CDW order in the vortex core until $t'\lesssim 0.5t$. For larger $t'$ values, we find a normal core with no CDW correlations.  
Figs.~2(a) and 2(b) show the profiles of superconducting and CDW order at selected values of $t^{\prime}$. 
The lower panels of Fig.~2 show the spatial maps of SC and CDW order parameters around a single vortex for $t^{\prime} = 0.3t$: CDW correlations can be clearly seen in the vortex core region. The same information is presented in spin language in Fig.~\ref{fig.SO3}(right). 

The SC and CDW profiles are, in fact, set by the same length scale, $\xi$, as the order parameters satisfy the uniform spin-length constraint. We obtain $\xi$ by fitting the SC profile to $\Delta(x)\sim \Delta_0\tanh(x/\xi)$; this functional form is consistent with the free energy in Eq.~\ref{eq.LG}\cite{Arovas1997}.  The resulting $\xi$ is plotted in the inset to Fig.~2(a). 
Separately, we define two length scales, $L_\Delta$ and $L_\phi$, as the full-widths at half-maximum of SC and CDW profiles respectively. We find that $L_\phi$ is always larger than $L_\Delta$ as shown in the inset to Fig.~2(b), although there is only one underlying length scale, $\xi$. Superconductivity is suppressed in the vortex core out to a radius set by $L_\Delta$. However, each vortex also hosts CDW correlations which extend to a larger distance, $L_\phi$.
This sets the stage for a coexistence phase, which can be understood as follows.
As the field is increased, the vortices are packed more and more tightly. Na\"ively, superconductivity persists until the inter-vortex distance approaches $L_\Delta$. Much before this, when the inter-vortex distance reaches $L_\phi$, the CDW regions around each vortex overlap. CDW order percolates throughout the system on top of the SC background, leading to a field-driven coexistence phase as we demonstrate below. Similar arguments have recently been put forward in the cuprates based on NMR results\cite{Wu2013}.

\begin{figure}[t]
\begin{center}
 \includegraphics[width=8.5cm]{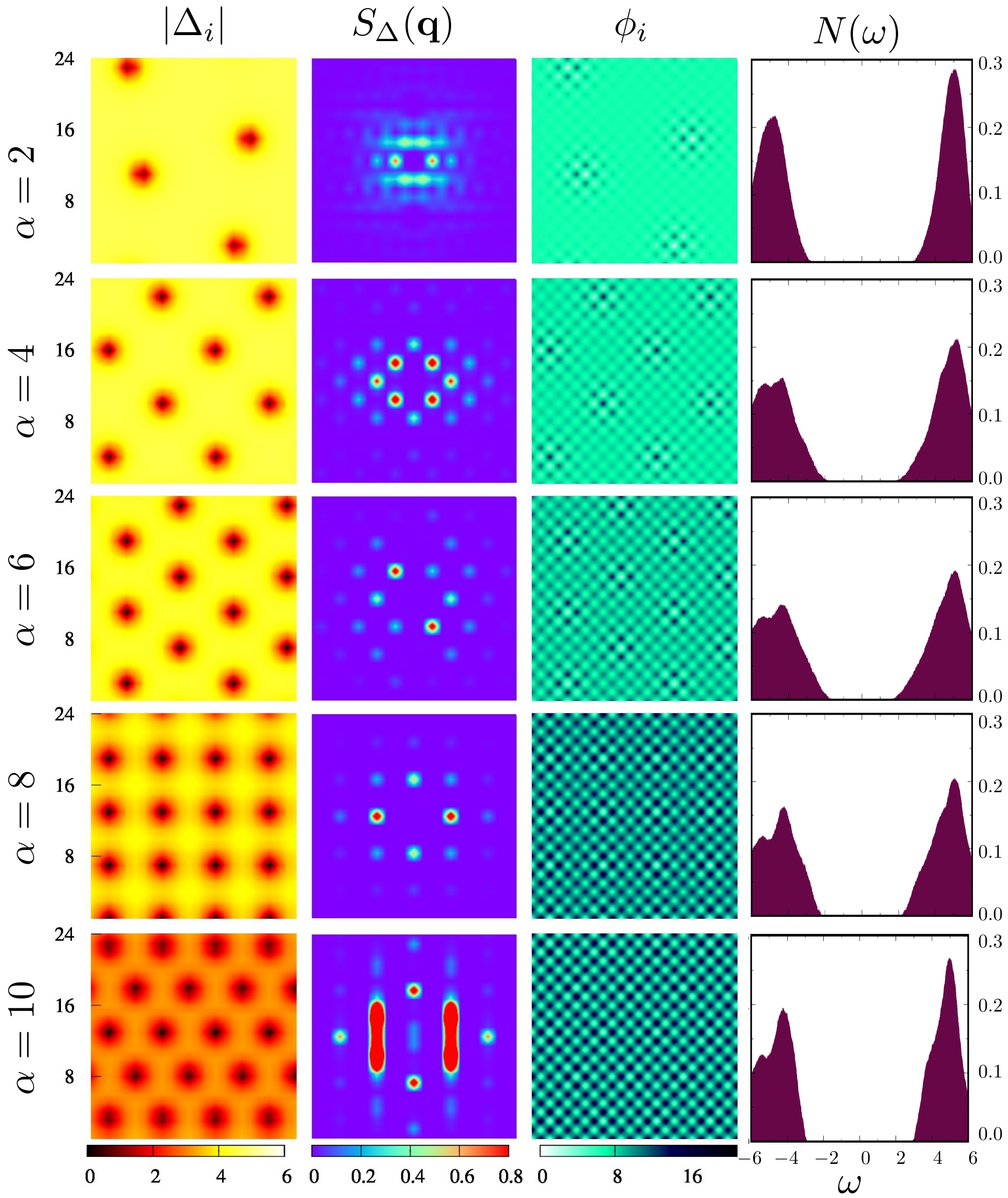}
\end{center}
\caption{Spatial maps of the SC order parameter amplitude, Fourier transform of 
SC amplitude and the density order parameter. The results are for $L=24$, $t^{\prime} = 0.2t$ and five different magnetic field values, parametrized by $\alpha$. We also show the density of states of fermionic excitations, $N(\omega)$, as a function of field.}
\end{figure}

\paragraph{Vortex lattice evolution:}
Fig.~3 shows our results for $t' = 0.2t$ with varying $\alpha$ (the total magnetic flux being $\alpha h/e$) on a $24\times24$ lattice. The panels show real space maps of $\Delta_{i}$ and $\phi_{i}$, showing the evolution of a vortex lattice with increasing flux. 
In addition, we plot the Fourier transform of the SC order parameter, defined as
$S_{\Delta}({\bf q}) = (1/N)\sum_{i}\vert \Delta_{i} \vert^{2} e^{i\bf q \cdot \bf{r}_i}$. 
The distribution of peaks in $S_{\Delta}({\bf q})$ reveals the geometry of the vortex lattice. 
The figure also plots the electronic density of states in the mean field ground state. 

For small fields, with $\alpha=2,4$, we find well separated vortices forming an anisotropic triangular lattice.  
At $\alpha = 6$, the lattice becomes near-isotropic. Upon increasing the field to $\alpha=8$, the vortices form a square lattice.  
This suggests an underlying phase transition driven by tuning vortex density. A similar transition into a second vortex lattice phase has been suggested in YBCO from torque magnetometry results\cite{Yu2016}.  
For $\alpha >8$, we find phase separation into square and triangular vortex lattices.
Finally, at $\alpha = 12$, we find a first order phase transition to a pure CDW phase, which has lower energy than solutions with SC order. Thus, at mean-field level, $H_{c2}$ is set by the competing CDW phase, unlike in conventional superconductors. 
\begin{figure}
 \begin{center}
  \includegraphics[width=3.23in]{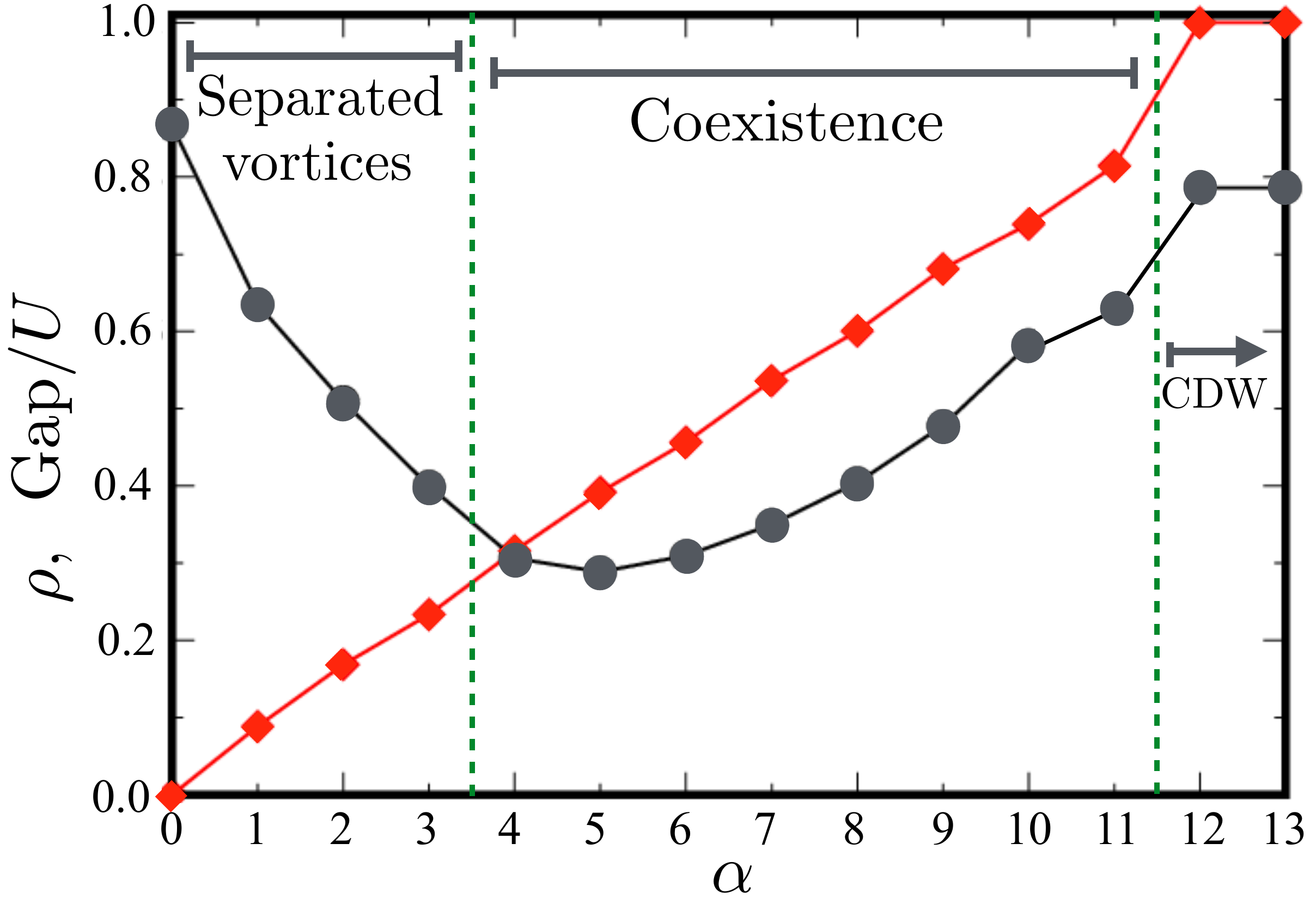}
 \end{center}
 \caption{CDW parameter $\rho$ (red diamonds) and the quasiparticle gap (black circles) as a function of flux $\alpha$ at $t^{\prime} = 0.2t$ as obtained from mean-field theory.
The CDW regions around vortex cores begin to overlap at $\rho \approx 0.3$, heralding coexistence.
The gap starts to increase beyond this threshold. In the high field `CDW' region, a pure CDW mean field state is favoured over a SC phase. 
}
\end{figure}

\paragraph{Phase coexistence:}

As seen in Fig.~3, every vortex core nucleates CDW correlations which begin to overlap when $\alpha=4$.  
The CDW order becomes progressively stronger with increasing field as vortex cores overlap more and more.  
For $\alpha \gtrsim 8$, we have near-uniform CDW order. 
With increasing field, the SC order weakens while the CDW order parameter grows. As a result, the electronic gap never closes, as shown in Fig.~3. 

Fig.~4 shows the in-field phase diagram for $t'=0.2t$. As we force a magnetic flux through Peierl's substitution, there is no $H_{c1}$ in our simulations.
To quantify the strength of CDW order, we define the parameter $\rho = \rho_{(\pi,\pi)}/\rho_{(0,0)}$ where $\rho_{\bf q}$ is the Fourier component of the density order parameter: $\rho_{\bf q} = \sum_i \phi_i e^{i{\bf q}\cdot {\bf r}_i }$. 
In a pure CDW state with site occupation oscillating between 0 and 2, $\rho$ takes the value unity. 
For small $\alpha$ values, we find $\rho$ to be small, indicating weak CDW order arising from well separated vortex cores. With increasing $\alpha$, $\rho$ increases monotonically. 
Within our mean-field theory, we find that CDW correlations begin to span the system when $\rho \sim 0.3$.
Based on this observation, we use $\rho \gtrsim 0.3$ as a heuristic criterion to signal macroscopic CDW order. 

While CDW and SC compete spatially, they both serve to open an electronic gap. At zero magnetic field, we have a uniform SC state with a large gap of the order of $U$. With an applied field, we induce vortices with CDW correlations, leading to a spatially textured $SO(3)$ order parameter field. 
Initially, for small magnetic fields, the order parameter gradients reduce the electronic gap. Once CDW order percolates throughout the system, the CDW order parameter no longer suffers sharp gradients and strengthens the gap once again. 

\paragraph{Discussion:}
Working in the context of the attractive Hubbard model, we have shown that competing CDW order emerges in vortex cores. At large fields, vortex cores overlap leading to a coexistence phase with SC and CDW correlations spanning the system -- a lattice version of a supersolid. The existence of a supersolid has been heavily debated in the context of liquid He\cite{Boninsegni2012}. Our study suggests that superconductors with competing phases are strong candidates for supersolidity.  

Our mean field theory in the strong coupling limit can also be seen as a spin problem with chiral interactions introduced by the orbital magnetic field. The vortex lattice ground state can be interpreted as a `meron crystal' -- a pseudospin state with a chiral texture\cite{Volovik2012}. When CDW order spans the system, this state spontaneously breaks a $\mathbb{Z}_2$ symmetry corresponding to two possible checkerboard density patterns, or equivalently to the choice between $z$ and $-z$ ordering of the spins. Our estimate for the superfluid stiffness (see Supplementary Material) indicates that this state is stable to fluctuations at intermediate fields. 
At large magnetic fields, the CDW order becomes much stronger than SC. In this regime, fluctuations may destabilize SC while leaving the CDW order intact. This suggests a `vortex liquid' phase with remnant CDW order and vanishing superfluid stiffness. This is an exciting direction for future study. In fact, field-induced coexistence and a charge-ordered vortex-liquid state have been reported in YBCO\cite{LeBoeuf2013,Gerber2015,Yu2016}.

We present a schematic phase diagram for the Hubbard model in the $t'$-$\alpha$ plane in Fig.~5. 
The plot shows the $t'< t_{c}^{'}$ region, where  
 $t_{c}^{\prime} \sim 0.5t$ is   
 the critical value beyond which CDW order is energetically unfavourable 
 at all magnetic fields. This phase diagram provides a reference point for understanding phase competition in the cuprates. For example, we find that SC is lost at $H_{c2}$ via a first order transition to the competing CDW phase; this may have been seen in YBCO using thermal conductivity measurements\cite{Grissonnanche2014}. 
We see that vortices themselves show interesting ordering phenomena with the vortex lattice changing from triangular to square configuration. This resembles the suggestion of two vortex solid states in YBCO from torque magnetometry measurements\cite{Yu2016}.

\begin{figure}
 \begin{center}
 \includegraphics[width=2.7in]{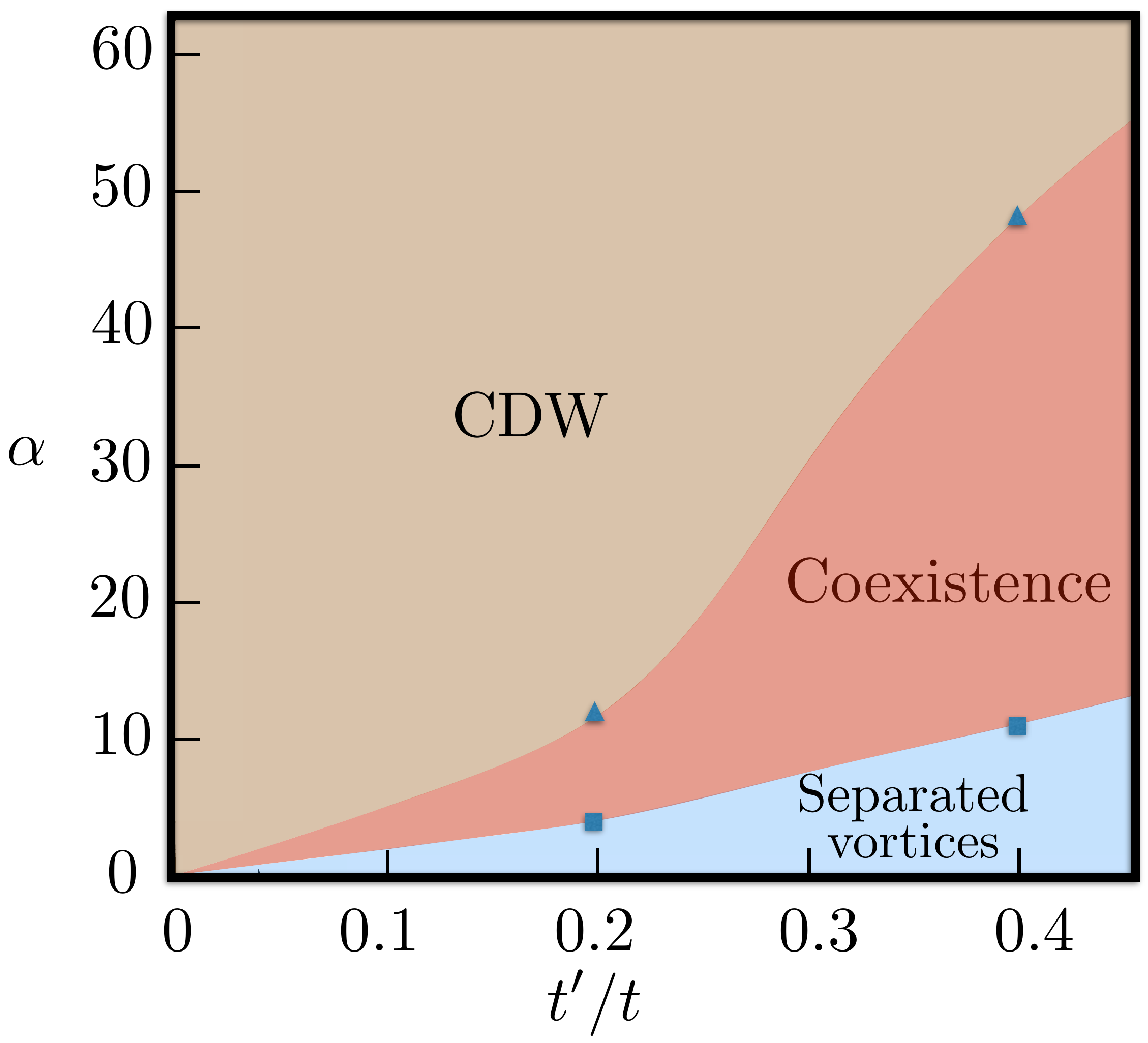}
 \end{center}
 \caption{Schematic phase diagram at $T = 0$ in 
 the $t^{\prime}-\alpha$ plane. 
The lower phase boundary represents a crossover field at which CDW correlations begin to span the system. The upper boundary represents $H_{c2}$, a first order phase transition into a pure CDW state.}  
 \end{figure}

While our study focusses on a simple model Hamiltonian, our results broadly apply to several material families which host competing orders. We have used $t'$ as a convenient handle to tune phase competition, this role could be played by experimentally tunable parameters such as doping in the cuprates \cite{forgan}, pressure in TiSe$_2$\cite{Kusmartseva2009}, etc.
Our results provide a theoretical paradigm to understand phase competition in these systems. \\
\textit{Acknowledgments}: We thank Arun Paramekanti, David Hawthorn, David Broun and Anton Burkov for discussions. 
MK acknowledges the use of High performance computing facility at Harish Chandra Research Institute, Allahabad, India.

\bibliographystyle{apsrev4-1}
\bibliography{HubbardSO3}

\newpage 
\section{SUPPLEMENTARY MATERIAL}
\renewcommand{\theequation}{S\arabic{equation}}
\renewcommand{\thefigure}{S\arabic{figure}}
\setcounter{equation}{0}
\setcounter{figure}{0}
\subsection{Peierl's substitution}
We have two types of hopping terms in our Hamiltonian: nearest-neighour hopping with amplitude $t$ and next-nearest hopping with amplitude $t'$. The phase of each hopping element represents a line integral of the vector potential, in accordance with the principle of Peierl's substitution. As the vector potential is not uniquely defined, there are several possible ways to assign the complex phases. The gauge invariant quantity is the magnetic flux: the sum of the hopping-phases along closed loops on the lattice. 

\begin{figure}
\includegraphics[width=3.25in]{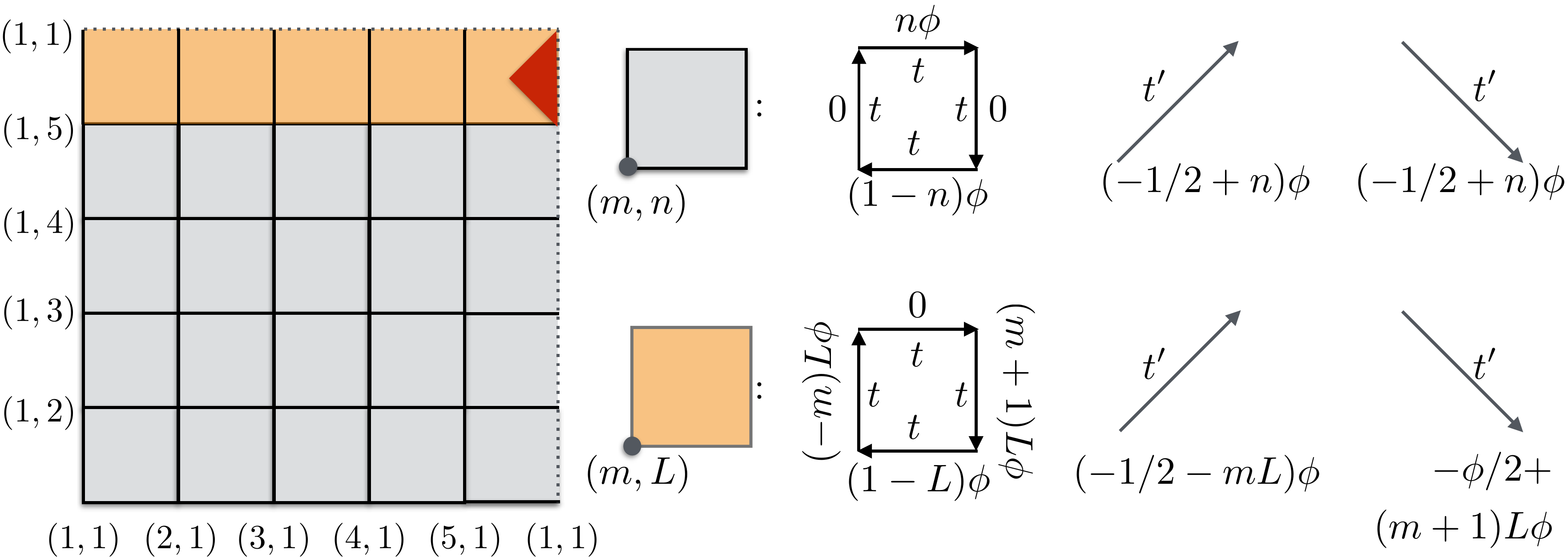}
\caption{Peierls substitution scheme on the $L\times L$ lattice. Left: The periodic cluster for $L=5$ for illustration purposes. Sites are labelled as $(m,n)$ -- not all site labels are shown. The bonds present in the cluster are depicted using dark solid lines. Bonds repeated due to periodic boundary conditions are shown in dotted lines. The square plaquettes are divided into two types as indicated by the colour. Right: Representative plaquettes of each type are shown, with the lower-left site labelled. The hopping-phases for the bonds on these plaquettes are shown. The parameter $\phi$ determines the flux through each square plaquette, given by $\hbar\phi/e$.  The Dirac string passes through the red triangular area. Due to the Dirac string, the flux is constrained to satisfy $\phi = 2\alpha\pi/L^2$, where $\alpha$ is an integer. }
\label{fig.Peierls}
\end{figure}

In our periodic $L\times L$ lattice, we assign hopping-phases so as to obtain a uniform magnetic flux. We use the scheme shown in Fig.~\ref{fig.Peierls}. We have introduced a parameter $\phi$ which encodes the phase picked up by an electron when hopping around any square plaquette, i.e., the magnetic flux through each square plaquette is $\hbar \phi/e$. Our square lattice system with periodic boundary conditions is equivalent to a torus, a closed surface. As shown by Dirac, the magnetic flux through any closed surface must be quantized in units of $h/e$ so that $\phi = 2\alpha \pi/L^2$, with $\alpha$ being an integer. The parameter $\alpha$ determines the total flux through the system, given by $\alpha h/e$. 
 
Dirac further argued that in the presence of a non-zero flux, we cannot define electronic wavefunctions on the surface smoothly.  The phase of the wavefunctions must wind around a singularity, which is called the `Dirac string'. In our scheme, the Dirac string passes through the red region within the top-right square plaquette in Fig.~\ref{fig.Peierls}. The sum of the hopping-phases around a contour which encloses this region has an additional contribution of $L^2 \phi = 2\pi \alpha$. This does not indicate an increased magnetic flux through this region; rather, it reflects a singularity in the definition of the electronic wavefunctions. Indeed, as this phase is a multiple of $2\pi$, it does not lead to any observable consequences. 

\subsection{Bogoliubov-de-Gennes formalism}

\begin{figure}
 \includegraphics[width=3.4in]{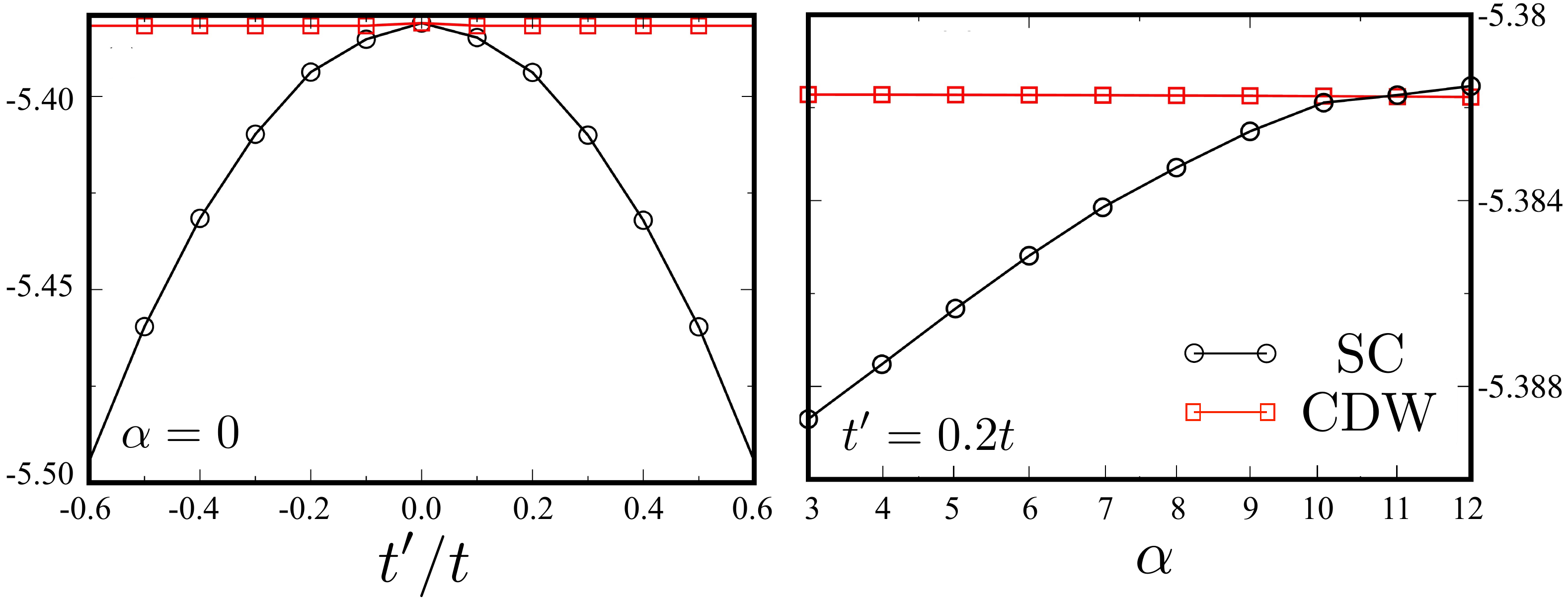}
\vspace{-0.15cm} 
\caption{Phase competition between SC and CDW orders. 
 (Left) Comparison of ground state energies as a function of $t'/t$, with the magnetic field turned off ($\alpha = 0$). At $t'=0$, the two orders are degenerate. A non-zero value of $t'$ lowers the energy of the SC state. (Right) Ground state energies as a function of magnetic flux, $\alpha$, with $t'$ fixed at $0.2 t$. At $\alpha =12$, the CDW state wins over the SC state indicating a first order phase transition. 
All energies are calculated at half-filling. }
 \label{fig.phasecompetition}
\end{figure}

The Hamiltonian for the Hubbard model is given in the main text. We decompose the on-site interaction term in pairing and density channels via a mean field 
decomposition. The complex superconducting order parameter is defined as, 
$\Delta_{i} = U\langle c_{i\downarrow}c_{i\uparrow}\rangle$, while the charge order parameter is defined as, 
$\phi_{i} = \frac{U}{2}(n_{i\uparrow}+n_{i\downarrow}) = 
\frac{U}{2}(\langle c_{i\uparrow}^{\dagger}c_{i\uparrow}\rangle + 
\langle c_{i\downarrow}^{\dagger}c_{i\downarrow}\rangle)$. The resulting effective Hamiltonian is given by
\begin{eqnarray}
\nonumber H_{MFT} & = & -t\sum_{\langle ij \rangle, \sigma}e^{i\theta_{ij}}c_{i\sigma}^{\dagger}c_{j\sigma}
 -t'\sum_{\langle\langle ij \rangle\rangle, \sigma}e^{i\chi_{ij}}c_{i\sigma}^{\dagger}c_{j\sigma}
 + h. c \\
  &-&\sum_{i,\sigma} \{ \mu + \phi_i \} c_{i\sigma}^{\dagger} c_{i\sigma} \nonumber 
 - \sum_{i}(\Delta_{i}c_{i\uparrow}^{\dagger}c_{i\downarrow}^{\dagger} + \Delta_{i}^{*}c_{i\uparrow}c_{i\downarrow}) \\
 &+& \sum_{i} \{\vert \Delta_{i}\vert^2 + \phi_i^2\}/{\vert U \vert}.
\end{eqnarray}
The hopping phases $\theta_{ij}$ and $\chi_{ij}$ are assigned according to the Peierl's substitution scheme described above.
We diagonalize this Hamiltonian using a Bogoliubov-Valatin transformation 
given by $c_{i\sigma}  =  \sum_{m}(u_{m i \sigma}\gamma_{m \sigma} - 
s_{\sigma}v_{m i \sigma}^{*}\gamma_{m, -\sigma}^{\dagger})$, where $\gamma_{m \sigma}^{\dagger}$
($\gamma_{m\sigma}$) creates (annihilates) a quasiparticle with spin $\sigma$ with 
energy $\epsilon_{m}^{\sigma}$ and wavefunctions $u_{m i \sigma}$ and $v_{m i \sigma}$.
We have introduced a spin index $s_{\uparrow} = 1$ and $s_{\downarrow} = -1$.
The resulting gap and number equations are  
\begin{eqnarray}
 \Delta_{i} & = & U\sum_{m}\left\{v_{m i \downarrow}^{*}u_{m i \uparrow}f(\epsilon_{m\uparrow})
 + u_{m i \downarrow}^{*}v_{m i \uparrow}f(\epsilon_{m \downarrow})\right\}, \nonumber \\
 n_{i\uparrow} & = & \sum_{m}\left\{\vert u_{m i \uparrow}\vert^{2}f(\epsilon_{m \uparrow}) 
 + \vert v_{m i \uparrow}\vert^{2}f(\epsilon_{m\downarrow})\right\}, \nonumber  \\ 
 n_{i \downarrow} & = & \sum_{m}\left\{\vert u_{m i \downarrow}\vert^{2}(1 - f(\epsilon_{m\uparrow}))
 + \vert v_{m i \downarrow}\vert^{2}(1 - f(\epsilon_{m \downarrow}))\right\},\nonumber\\
\end{eqnarray}
where $f(\epsilon_m) = 1(0)$ if $\epsilon_{m}<0(>0)$ is the Fermi function at zero temperature.  
Starting from initial guess values, we iterate these equations to obtain self-consistent values of $\Delta_i$ and $\phi_i$. The chemical potential is tuned to fix the density at half-filling. \begin{figure*}
 \includegraphics[width=5in]{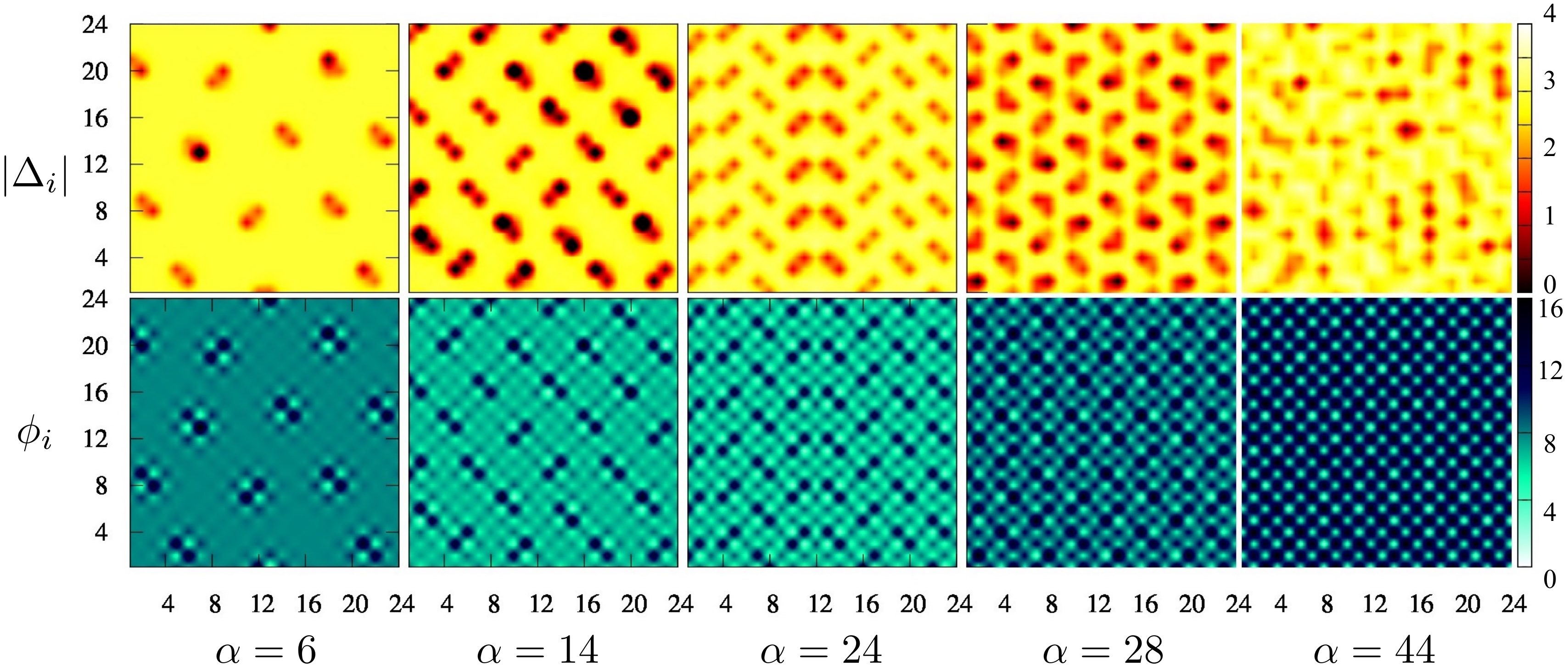}
 \caption{Spatial maps of the pairing amplitude ($\vert \Delta_{i}\vert$) 
 and the CDW ($\phi_{i}$) order  
 with changing magnetic field at $t^{\prime} = 0.4t$ on a 24$\times$24 lattice.}
\label{fig.tp4}
\end{figure*}

\subsection{Phase competition}

For a given set of parameters $t'$, $U$ and $\alpha$, we find several self-consistent solutions. 
In particular, we find a pure CDW state with $\Delta_i = 0$.  
To illustrate the phase competition in the Hubbard model, we compare the energy of this CDW state with that of the SC state in 
Fig.~\ref{fig.phasecompetition}. In the absence of a magnetic field ($\alpha=0$), the two states are degenerate when $t'=0$, 
while a non-zero $t'$ lowers the energy of the SC state. 

Competition between orders can also be tuned by increasing the magnetic field. This is depicted in Fig.~\ref{fig.phasecompetition}(right). When $\alpha$ is increased at fixed $t'$, 
the energy of the CDW state does not change as the CDW state is an insulator. However, in the SC phase, increasing $\alpha$ introduces more vortices and increases the energy.
The energy of the SC state steadily rises and eventually crosses the CDW energy. This signals $H_{c2}$ at the mean-field level, with CDW order becoming energetically favourable over a superconducting vortex state.

\subsection{Higher $t^{\prime}$ regime}

With increasing $t'$, the radius of the CDW region in each vortex core shrinks. 
Consequently, the threshold field for coexistence increases.  
Fig.~\ref{fig.tp4} shows spatial maps of the SC and CDW order parameters for $t'=0.4t$ which 
can be compared with the $t'=0.2t$ data in Fig.~3 of the main text. 
The percolation of CDW correlations is slow to occur with a coexistence state only setting at $\alpha\sim11$.
Finally, $H_{c2}$ is encountered at $\alpha\sim48$, when the CDW state becomes energetically favourable. 
Unlike the case of $t'=0.2$ discussed in the main text, it is difficult to discern changes in the geometry of the vortex lattice here due to the high density of vortices.

\subsection{Superfluid stiffness}

\begin{figure}
 \begin{center}
\includegraphics[width=3.2in]{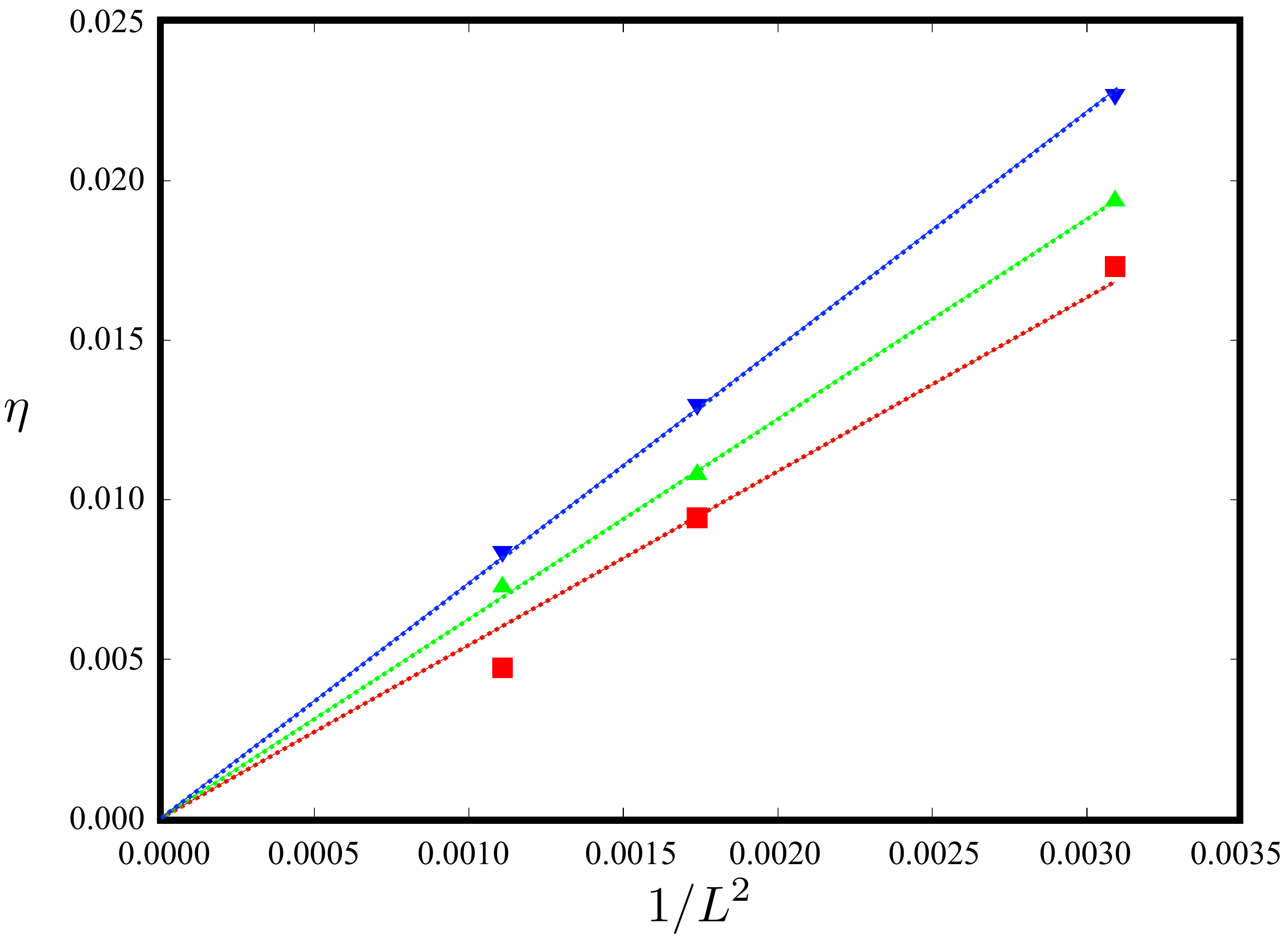}
\end{center}
\caption{Superfluid stiffness. We plot $\eta(L)$ vs. $1/L^2$ for three sets of $\{\alpha,L\}$ values: (a) $(\{0,18\},\{0,24\},\{0,30\})$ (blue downward triangles), (b) $(\{2,18\},\{4,24\},\{6,30\})$ (green upward triangles), and (c) $(\{4,18\},\{8,24\},\{12,30\})$ (red squares). The lines are fits to the form $\eta = \rho/L^2$. All three best-fit lines have positive slopes indicating a positive superfluid stiffness. }
\label{fig.stiffness}
\end{figure}
Our mean-field results indicate coexistence of SC and CDW orders, forming a supersolid state. To check if this phase is stable to fluctuations, the standard diagnostic is superfluid stiffness -- which measures the energy cost of imposing a smooth gradient in the SC order parameter. 
To estimate the stiffness, we take the following route. We introduce an additional component of the vector potential $\mathbf{A}_{tangential} = 2\pi\hat{x}/L $. 
If we were to turn off the orbital magnetic field, this vector potential leads to a `flowing' superfluid solution with $\Delta_i \sim \Delta_0 e^{i4\pi x_i/L} $. The resulting energy cost is a measure of superfluid stiffness. For a simple superfluid with no competing order, this energy cost (the increase in energy per site) is proportional to $\rho/L^2$, where $\rho$ is the superfluid stiffness and $L$ is the system size. We define $\eta(L) = E_{2\pi} - E_0$, where $E_{2\pi}$ is the energy (per site) of the flowing state. This is calculated by using the $\Delta_{i}$ values obtained after inclusion of the tangential vector potential to evaluate the expectation value of the mean-field Hamiltonian. 

The obtained $\eta(L)$ values are plotted as a function of $1/L^2$ in Fig.~\ref{fig.stiffness}. In the $\alpha=0$ case (no orbital magnetic field), we see that $\eta(L)$ indeed scales as $1/L^2$, with a positive slope. This slope is proportional to the superfluid stiffness.   

In the presence of the orbital field, we seek to plot $\eta(L)$ for configurations with the same magnetic flux density across different system sizes. In our calculations, the magnetic flux density is $\alpha h/eL^2$, with $\alpha$ being an integer and $L$ ranging from $18-30$ (for smaller sizes, we see strong finite size effects). Generically, it is not possible to find multiple $\{\alpha,L\}$ values for which $\alpha/L^2$ is a constant. In Fig.~\ref{fig.stiffness}, we plot $\eta(L)$ for $\{\alpha,L\} = (\{2,18\},\{4,24\},\{6,30\})$ and $(\{4,18\},\{8,24\},\{12,30\})$ which correspond to approximately constant $\alpha/L^2$ values. The resulting $\eta(L)$ values also scale linearly with $1/L^2$ with a positive slope. We conclude that these superfluid stiffness is positive for these flux densities. We also note that the stiffness decreases with increasing flux density. In particular, we note that the stiffness is positive for the flux densities corresponding to $\alpha = 4$ and $\alpha=8$ on a 24$\times$24 lattice. As discussed in the main text, these parameters have macroscopic CDW order with CDW correlations spanning the entire system. This suggests that the coexistence phase is stable to fluctuations. 

It is possible that that stiffness may vanish at higher flux densities, perhaps close to $H_{c2}$. In this regime, the CDW order parameter becomes approximately constant while the SC order parameter suffers large gradients due to the presence of vortices.
It is then conceivable that fluctuations can wash out the in-plane order while preserving  order in the $z$-direction. This will lead to a `pairing liquid' state (analogous to a spin liquid) with remnant CDW ordering. Equivalently, this can be understood as melting of the vortex lattice. Below this threshold, the inter-vortex distances are fixed by strong interactions between vortices. A small amount of disorder will then suffice to pin the entire vortex lattice so as to generate a robust SC state. However, when fluctuations wash out the coherence in the SC, the vortices become mobile giving rise to a `vortex liquid'. This is an interesting direction for future study.

\end{document}